\documentclass[draft]{agujournal2019}
\usepackage{url} 
\usepackage{lineno}
\usepackage[inline]{trackchanges} 
\usepackage{soul}
\usepackage{amsmath}
\draftfalse

\journalname{JGR: Solid Earth}

\begin{document}

%
%


\title{Receiver Functions in the San Fernando Valley, California: Graph-Regularized Bayesian Approach for Gravity-Informed Mapping}

%
%




\authors{Valeria Villa\affil{1}, Robert W. Clayton\affil{1}, Patricia Persaud\affil{2}}
\affiliation{1}{Seismological Laboratory, California Institute of Technology, Pasadena, California, USA}
\affiliation{2}{Department of Geosciences, University of Arizona, Tucson, Arizona, USA}






\correspondingauthor{Valeria Villa}{vvilla@caltech.edu}




\begin{keypoints}

\item A Bayesian framework to interpret receiver functions is augmented with gravity data and a connecting graph to neighboring stations' data.
\item Analysis of a synthetic dataset can resolve two different densities and the interconnecting gradient.
\item In the San Fernando Valley, the Sylmar sub-basin is  5.6 km deep, and the San Fernando sub-basin is 4 km.
\end{keypoints}

%
%

%
%


\begin{abstract}
The San Fernando Valley (SFV) in Southern California is a complex sedimentary basin whose shape strongly influences ground shaking. We develop a fully quantitative, probabilistic graph-regularized inference model that integrates both gravity and receiver function (RF) constraints and evaluate its ability to determine the basin's shape. The sediment-basement interface in single-station RFs is often difficult to interpret due to scattering and noise, which can render isolated stations unusable. By using RFs from a dense seismic array and incorporating gravity, we address the issue of non-uniqueness in converting the times of RF phases to layer thickness by comparing the predicted gravity to observations at each station. In areas where the density contrast may change, Bayesian inference with a graph Laplacian allows us to determine the effective density contrast by taking into account its neighbors’ picks and densities.  This method promotes spatial smoothness between neighboring stations, while preserving sharp contrasts in locations supported by the RF and gravity data. We applied this method to a dataset that was acquired in fall 2023, when 140 nodes were installed in the SFV. Our results show the deep Sylmar sub-basin, the San Fernando sub-basin, and the Leadwell high found in a previous study (Juárez‐Zúñiga and Persaud, 2025), and our results also show good agreement with the industry seismic reflection profiles across the valley. This method demonstrates how to incorporate gravity with lateral density variations into receiver function interpretation to better map interfaces in the subsurface.
\end{abstract}

\section*{Plain Language Summary}

Understanding seismic hazard in densely populated urban areas is a subject of great interest and importance to society, but it involves careful understanding of the subsurface structure. Receiver functions (RF) are a tool used to map subsurface layers, but in the past, have mostly been interpreted through heuristic, non-quantitative methods. Here, we introduce a method, based on Bayesian statistics, to constrain the receiver function interpretation. The essence of this method is in taking into account gravity measurements, as well as RF measurements from neighboring stations. To show our method's application, we used it on synthetic data and a new dataset recorded by seismic nodes deployed across the SFV. Our results show good agreement with previous studies that used independent techniques. This demonstrates the success of modern mathematical frameworks for studying sedimentary basins, which can then inform seismic hazard models.

%
%

\section{Introduction}
The San Fernando Valley (SFV) in Southern California is a densely populated urban basin with a population of 1.8 million \cite{us_census_bureau_san_nodate}. Its tectonic setting as a sedimentary basin heightens seismic hazard by trapping and amplifying seismic waves \cite{bard_two-dimensional_1985}. This hazard is compounded by the numerous active fault zones that bound and cross the valley, including those responsible for the destructive 1971 Mw 6.7 San Fernando and 1994 Mw 6.7 Northridge earthquakes \cite{palmer_san_1971, hough_17_2024}, and newly identified active structures in the southern part of the valley \cite{omojola_detecting_2025}. Understanding the basin’s shape is therefore important as its deep and irregular geometry strongly amplifies and prolongs shaking \cite{bonilla_site_1997}, implying that current models may underestimate the true hazard \cite{clayton_exposing_2019}. Here, we investigate the depth and structure of the SFV basin, using a new method to provide constraints on this key control of seismic hazard.

\citeA{langenheim_preliminary_2000} mapped the SFV and suggested the presence of a deep ($> 5 $ km) basin in its northern part. A later study, \citeA{langenheim_structure_2011} , further elucidated this and confirmed other important layers in the basin, concluding the deepest part of the basin could range between 5--8 km. These studies used industry reflection profiles, gravity, aeromagnetic data, and boreholes to arrive at these conclusions. More recently, a new dense seismic array study imaged the basin depth by combining their 3D velocity model from ambient noise tomography with gravity inversion and integrating horizontal-to-vertical spectral ratios and aeromagnetic data \cite{juarezzuniga_new_2025}. That study concluded that the valley consists of two sub-basins: the San Fernando Basin and the Sylmar Basin. The Sylmar Basin was found to be deeper, reaching a depth of 6.5 km, while the San Fernando Basin extends to about 4 km depth near the Northridge Hills fault. However, RFs have not yet been computed for this dataset, and this is one of the goals of this paper. 

RFs are a well-established method for estimating interface depths based on P-to-S converted phases at subsurface boundaries. Originally established as a routine method in the 1980s for investigating interfaces such as the Mohorovičić discontinuity (Moho), they have since become a standard tool for imaging various crustal structures \cite{lawrence_global_2006, piana_agostinetti_sedimentary_2019, ramirez_moho_2021, esteve_mapping_2025}. However, applying the RF method in urban settings has not generally been done because the results can be difficult to interpret due to high levels of noise and scattering, depending on the data quality. One of the first successful urban applications of this approach was largely enabled by a dense linear array of broadband stations, which allowed the PpPs phases related to the P-to-S conversions to appear coherently along the array and be distinguishable \cite{ma_structure_2016}. More recently, using ten dense linear arrays of seismic nodes, the P-to-S phase from the basin bottom could be observed and mapped across the San Gabriel, Chino and San Bernardino basins in Southern California \cite{ghose_basin_2023, liu_structure_2018, wang_urban_2021}.

Previous studies have worked on ehancing the interpretability of RFs. The most elementary method employed is stacking RFs from multiple teleseismic events, to increase the signal-to-noise ratio of key conversions \cite{vinnik_detection_1977}. Through this method, clear phases were able to identify the Moho discontinuity beneath Southern California \cite{yan_regional_2007, ozakin_systematic_2015}. However, one drawback of this approach is its dependence on the number and quality of RFs that can be stacked. This poses a challenge for temporary deployments, where the short duration limits the number of recorded teleseismic events. In addition, in the Los Angeles basin, it was shown that using the permanent broadband stations to identify the P-to-S phase that corresponds to the sediment-basement interface is difficult and often not successful \cite{ma_structure_2016}.

Some studies have shown that semi-automated interpretation methods that incorporate additional constraints extending beyond traditional RFs analysis can be applied to image basin structure in urban environments. In the northern Los Angeles basins, \citeA{wang_urban_2021} used a Bayesian coherence function that incorporated information from neighboring stations, while \citeA{villa_three-dimensional_2023} combined gravity measurements with RFs to map the sediment–basement interface. Both approaches were successful, in large part because they were applied to a temporary nodal deployment with very dense linear arrays, with stations spaced only ~250 m apart. This geometry allowed phase coherence to be tracked across stations with high resolution, greatly enhancing the ability to identify the basin bottom. However, more recent nodal deployments have shifted away from solely dense linear arrays to include scattered station geometries, in order to have more uniform spatial coverage and meet additional scientific objectives beyond RFs analysis. 

Here, we present a probabilistic and quantitative method that incorporates gravity data to identify and refine P-to-S phase picks at single stations, enabling more reliable identification of the sediment-basement interface even under challenging observational conditions. We apply this method to data recorded by the nodal array deployed in 2023 in the SFV in Southern California. While previous subsurface datasets have been collected by the oil industry, they are very limited in coverage, and the most comprehensive results to date come from a study carried out with the new nodal dataset to produce 3D shear-wave velocity and basin depth models \cite{juarezzuniga_new_2025}. However, RF analysis, which offers one of the most direct measurements of basement depth and, in other studies, potential intra-crustal layers, has not yet been applied to this dataset.

\section{Data and Methods}

\begin{figure}[ht]
\centering
\includegraphics[width=\textwidth]{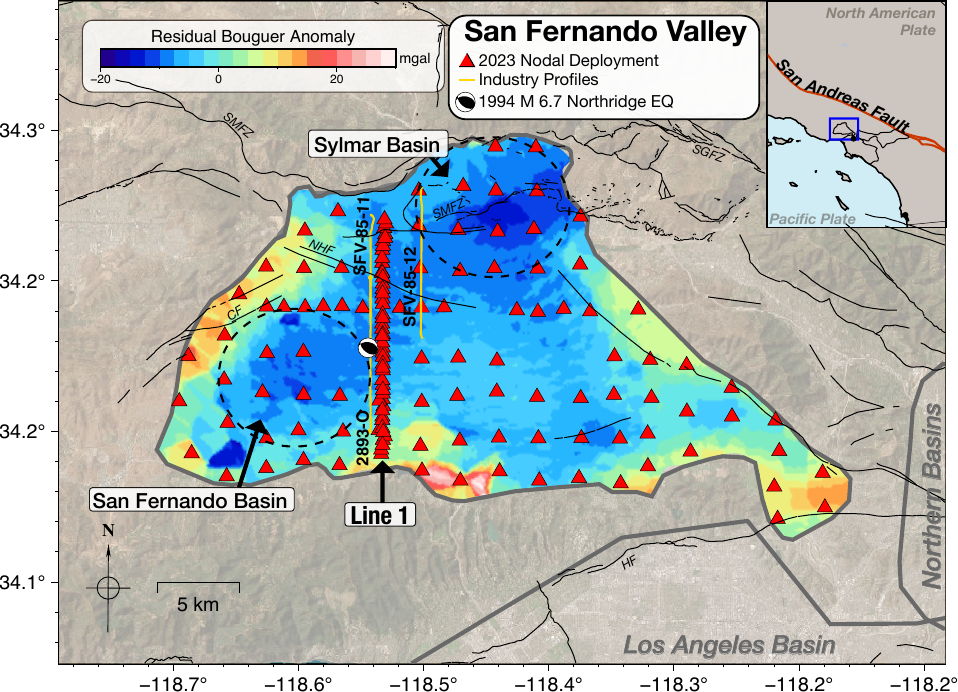}
\caption{Map of residual Bouguer gravity anomaly across the San Fernando Valley, California, with color scale indicating residual gravity values. Red triangles show node station locations. The Line 1 nodes crossing the epicentral region of the 1994 Mw 6.7 Northridge earthquake, are marked. Thin black lines represent Quaternary faults \cite{jennings_fault_2010}, while the thick gray line indicates the basin's boundary polygon from \citeA{juarezzuniga_new_2025}, and the black dashed lines mark the approximate sub-basins. The yellow lines show industry profiles from \citeA{langenheim_structure_2011}. The inset shows the region's location with a blue rectangle in relation to major tectonic features such as the Pacific and North American plate boundary. CF - Chatsworth Fault; HF - Hollywod Fault; NHF - Northridge Hills Fault; SGFZ -San Gabriel Fault Zone; SMFZ - Sierra Madre Fault Zone.}
\label{fig1_gravmap}
\end{figure}

In fall 2023, 140 short-period, three-component SmartSolo nodes were deployed by a group of 29 volunteers across the SFV (Fig. \ref{fig1_gravmap}; \citeNP{patricia_persaud_san_2023, persaud_volunteer-led_2024}). The array had an average spacing of 1.40 km, with one north–south line sampled more densely at 250 m intervals. The station geometry consisted of 49 stations along the north–south line, 15 along an east–west line, and the remaining stations distributed in a shotgun pattern (Figure \ref{fig1_gravmap}). The deployment lasted for approximately 30 days, with all stations recording continuously at a sampling rate of 500 Hz. We used the 10/31/23 Chile M6.7 teleseismic event with an epicentral distance of $76.8^\circ$ from the array to compute the RFs. The data were downsampled to 50 samples per second and windowed to 30 seconds around the P-wave onset. They were rotated to the ZRT coordinate system, and we applied standard iterative time-domain deconvolution \cite{ligorria_iterative_1999} using a Gaussian parameter of 2.5 to filter the RFs between ~0.1 and 1 Hz.

We obtained gravity data from the Pan-America Center for Earth and Environmental Sciences gravity portal \cite{paces_paces_2012}. We applied nearest-neighbor inverse-distance weighting to interpolate the Bouguer gravity points onto a 100-meter grid for the SFV study area (Fig. S1). Residual gravity, obtained by removing the regional trend, was used to emphasize local basin effects. We applied the same method used by \citeA{juarezzuniga_new_2025} to determine the residual Bouguer gravity, in which the regional anomaly is defined by convolving the Bouguer field with a 2D Gaussian kernel of 5 km half-width and then subtracting the regional anomaly from the Bouguer gravity anomaly to enhance short-wavelength, near-surface features. Figure \ref{fig1_gravmap} shows the residual Bouguer gravity within the SFV polygon boundary from \citeA{juarezzuniga_new_2025}. 

\subsection{Method}

\subsubsection{General Description}

\begin{figure}[ht]
\centering
\includegraphics[width=\textwidth]{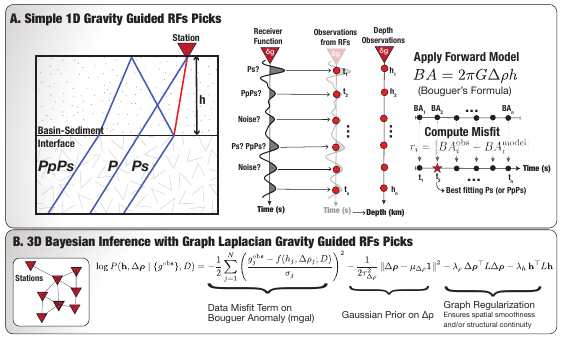}
\caption{A) Schematic illustrating the model used for gravity-guided receiver function interpretations. The diagram shows the integration of gravity constraints and P-to-S converted phases to estimate basin depth. (B) The model is then solved using GravNet, a Bayesian inference framework with graph Laplacian regularization, which enforces spatial smoothness by connecting neighboring stations through a network graph.}
\label{schematic_fig1}
\end{figure}

We introduce a method to combine gravity estimates with RFs to find the sediment-basement interface referred to as GravNet where we jointly estimate the density contrast and interface depth \textit{h} beneath a network of seismic stations. The inference is performed in a Bayesian framework where we sample the effective density contrast, $\Delta \rho$. In other words, the only free parameter is the effective density contrast $\Delta\rho$, from which depth $h$ and predicted residual Bouguer gravity are derived. Afterwards, time picks are selected based on the depth $h$. To maintain a coherent structural signal, we apply a graph Laplacian regularization term to both the density contrast and the subsurface depth \textit{h}. Without this graph, each station would be solved independently, often producing noisy or inconsistent parameter jumps due to local data ambiguity. The graph Laplacian couples neighboring stations, discouraging abrupt differences and encouraging the sampler to adopt values that remain compatible with neighboring picks. As a result, the inferred structure varies smoothly in space unless the data strongly support a real discontinuity. Although gravity is used to guide the selection of the best RF phase pick, the final output is derived directly from the RFs. Each local maximum in the RF of a station is considered as an observation, regardless of its amplitude. Because the basin consists of slow sediments overlying faster crystalline basement, the corresponding \textit{Ps} and \textit{PpPs} conversion is positive; therefore, we focus only on positive RF phases. Negative phases, which arise from velocity inversions or low-velocity zones, can also be incorporated if one is interested in those types of geological structures, but they are not the target of this study.

\subsubsection{Detailed Specifications}

The penalty function in the model is Bouguer's formula, \(\delta g = 2\pi G \Delta\rho h\), where $G$ is the gravitational constant, \(\Delta \rho = \rho_{sed} - \rho_{basement}\) is the density contrast between the sedimentary \(\rho_{sed}\) and basement rock \(\rho_{basement}\), and \(h\) is the thickness of the sedimentary layer. At each station, the observations from RFs are converted to depth using either a constant velocity or a velocity model. The conversion from time to depth is as follows for either phase, Ps or PpPs, where p is the ray parameter \cite{zhu_moho_2000}
\begin{equation}
h = \frac{t_{P_s}}{\sqrt{\frac{1}{V_s^2} - p^2} - \sqrt{\frac{1}{V_p^2} - p^2}}
\label{eq:Ps}
\end{equation}

\begin{equation}
h = \frac{t P_p P_s}{\sqrt{\frac{1}{v_s^2} - p^2} + \sqrt{\frac{1}{v_p^2} - p^2}}
\label{eq:PpPs}
\end{equation}
Therefore, the model has \(d_{model} = f(h,\Delta \rho)\) where \(f(h,\Delta \rho) = 2 \pi G \Delta\rho h\), and $d_{obs}$ is the possible basin depth values converted from time in the RFs using Equations \ref{eq:Ps} and \ref{eq:PpPs}. As a simple first-order estimate, we assume a constant $\Delta \rho$ across each station to evaluate the fit to the gravity and the corresponding receiver-function time pick. This step is easy to compute and provides a practical way to guide the most significant pick. This serves as a way to set the prior in the Bayesian framework.

However, we know that the density contrast may vary across a large region due to geologic changes, different rock types, etc. Thus, we invoke a graph-inference approach to solving this inference problem. The methodology is as follows.

Assuming Gaussian errors, the likelihood is given below.
\begin{equation}
    \mathcal{L}(\Delta \rho) = \frac{1}{\sqrt{2\pi}\sigma} \exp\left(-\frac{(d_{\text{obs}} - f(g, \Delta \rho))^2}{2\sigma^2}\right)
    \label{eq:likelihood_gaussian}
\end{equation}
As prior information, we assume that $\Delta \rho$ follows a uniform distribution. The thickness parameter $h$ is selected from a discrete set ${h_i}$ and we evaluate the likelihood for each $h_i$. The posterior sampling $P(A \mid B) \propto P(B \mid A)\, P(A)$, where \( P(A \mid B) \) is the posterior, \( P(B \mid A) \) is the likelihood, and \( P(A) \) is the prior.

This study employs graph-based regularization to ensure spatial smoothness across RFs. The regularization is handled algorithmically, making it particularly useful for receivers with irregularly distributed locations such as the shotgun stations shown in Figure \ref{fig1_gravmap}. Additionally, the method assigns a quantitative score to each pick, indicating the robustness of the resulting image.

The method works as follows. We let $\theta_j = \Delta \rho_j$ be the estimated density contrast at station $j$. To impose spatial coherence, we introduce a graph-based regularization term using the graph Laplacian $L$.
\[
\Phi_{\text{reg}} = \lambda \sum_{i,j} w_{ij} (\Delta \rho_i - \Delta \rho_j)^2 = \lambda \boldsymbol{\theta}^\top \mathbf{L} \boldsymbol{\theta}
\]
where \( \lambda \) is the regularization strength, \( w_{ij} \) is the weight between nodes \( i \) and \( j \), and \( \Delta \rho_i \) represents the model parameter (e.g., density contrast) at node \( i \). The right-hand expression is the quadratic form of the graph Laplacian \( \mathbf{L} \), with \( \boldsymbol{\theta} \) being the vector of parameters \( \Delta \rho \) across all nodes. We further include regularization on the thickness $h$,
\[
\Phi_{\text{reg}} = \lambda_\rho \sum_{i,j} w_{ij} (\Delta \rho_i - \Delta \rho_j)^2 + \lambda_h \sum_{i,j} w_{ij} (h_i - h_j)^2 = \lambda_\rho \boldsymbol{\theta}_\rho^\top \mathbf{L} \boldsymbol{\theta}_\rho + \lambda_h \boldsymbol{\theta}_h^\top \mathbf{L} \boldsymbol{\theta}_h
\]

\[
w_{ij} = \frac{1}{\| \mathbf{x}_i - \mathbf{x}_j \|}
\]

Here, \( w_{ij} \) represents the weight between node \( i \) and node \( j \), defined as the inverse of their Euclidean distance. This formulation ensures that nearby nodes exert stronger influence during regularization than distant ones. The weights are only computed for the \( k \)-nearest neighbors of each node, as specified by the user. That is, for a given node \( i \), \( w_{ij} = 0 \) unless node \( j \) is among the \( k \) closest nodes to \( i \). This localizes the regularization, promoting spatial smoothness only among physically close stations. 

To estimate the posterior probability of the density contrast \( \Delta \rho_j \) at node \( j \), we adopt a Bayesian formulation that includes two main components: (1) a likelihood term capturing the misfit between the observed data and the forward prediction, and (2) a graph-based regularization term that causes spatial smoothness of the density contrast \( \Delta \rho \) across neighboring nodes. Assuming a uniform prior over \( \Delta \rho_j \), the log-posterior simplifies to:
\[
\log P(\Delta \rho_j \mid d_j, h_j) = -\frac{1}{2} \left( \frac{d_j^{\text{obs}} - f(h_j, \Delta \rho_j)}{\sigma_j} \right)^2 
- \lambda_\rho \sum_{k \in N(j)} w_{jk} (\Delta \rho_j - \Delta \rho_k)^2
\]

The density regularization parameter $\lambda_\rho$ controls how similar the density contrast must be between neighboring stations, and the depth regularization parameter $\lambda_h$ plays the same role for the interface depths. Small values of $\lambda_\rho$ or $\lambda_h$ allow for more lateral variability, whereas large values make the corresponding field (density or depth) nearly uniform. The $k$-nearest-neighbor parameter defines how many nearby stations are linked in the graph and therefore how far this smoothing influence extends spatially. We note that these regularization parameters are fixed, and that, as is implied by $\Delta\rho$ being on the left-hand side of the bar in the above equation, the only free parameter is $\Delta \rho$.

For each fixed value of \( h_j^{(i)} \), we run \texttt{emcee}, a Markov Chain Monte Carlo (MCMC) ensemble sampler, to draw samples from the posterior distribution of \( \Delta \rho_j \) \cite{foreman-mackey_emcee_2019}. The likelihood includes a graph-based regularization term that incorporates the current values of neighboring stations, effectively serving as a spatial prior. For each trial \( h_j^{(i)} \), we store the full set of samples of \( \Delta \rho_j \), which include the data misfit and the graph-based regularization. The sampler is run for 25,000 steps, where the first 5,000 are taken as the ``burn-in"  period. At the end of the routine, we are left with samples from the posterior distribution of \( \Delta \rho_j \).  Posterior distributions for other parameters derived from $\Delta\rho$ are then generated from individual samples of $\Delta\rho$. Finally, median values and quantiles for each parameter are calculated from their individual posterior distributions.


%


%
%
%
%

\section{Results}
In this section, we first present the receiver-function results for the selected teleseismic event recorded by the SFV array. We then introduce the baseline gravity-guided approach, which assumes a constant density contrast at all stations. Finally, we present the Bayesian model, which incorporates receiver-function picks from neighboring stations and a prior on density. The advantages of the Bayesian approach over the baseline are further illustrated with a synthetic example.

\subsection{Receiver Functions}
\begin{figure}[ht!]
\centering
\includegraphics[width=\textwidth]{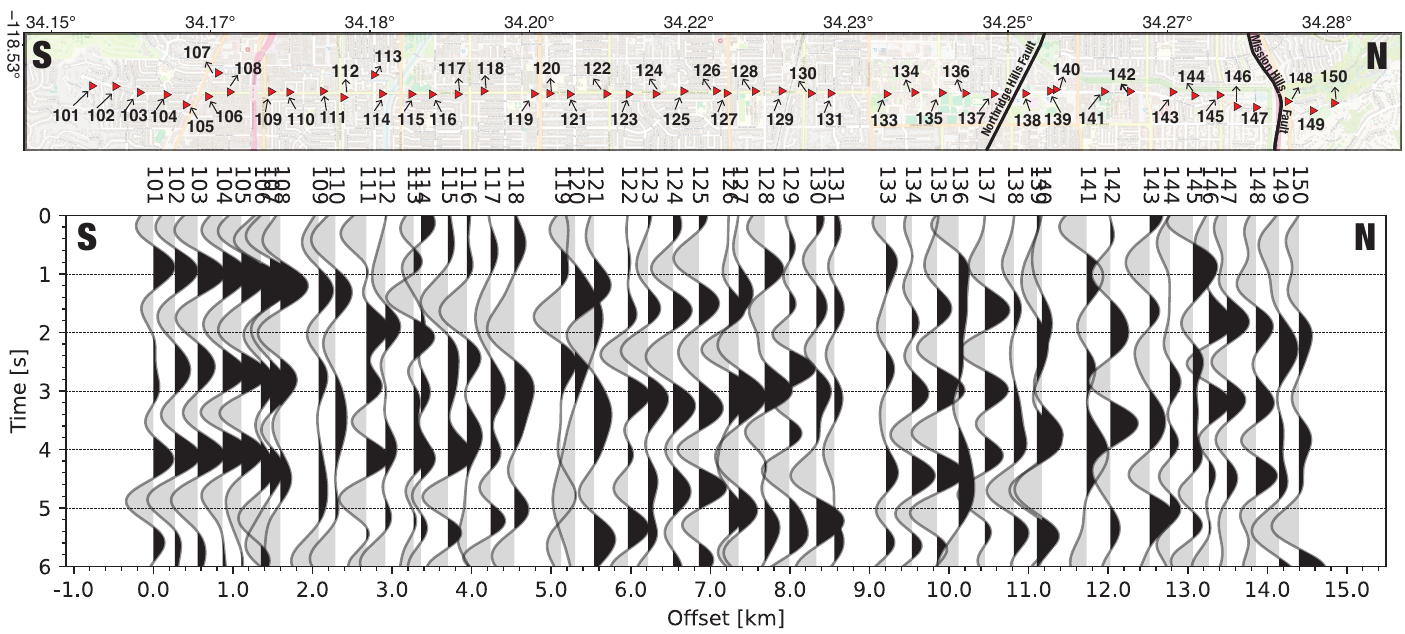}
\caption{1 Hz receiver functions for the 10/31/23 Chile M6.7 event recorded along Line 1, the south-to-north station profile of the SFV nodal array shown in Figure 1. The top panel shows the expanded map view of station locations, with faults indicated by black lines.}
\label{rfs_fig3}
\end{figure}

The SFV RFs provide an interpretable image primarily along Line 1 (Fig. \ref{rfs_fig3}), where most stations show 2–4 positive phases in the first 6 s of the RFs. In the southern part of the profile, phase coherence is clear and corresponds to the basin shallowing toward the foothills of the Santa Monica Mountains. Station 120 shows poor results due to elevated noise levels and reduced waveform coherence in the raw seismograms, particularly on the north component (Fig. S2–4), and is excluded from further analysis. From stations 133 to 150, phase coherence becomes more complex, likely reflecting a combination of reverberations in the basin, scattering, and anthropogenic effects. While usable RFs are obtained at stations outside of Line 1, their coherence is more difficult to interpret in isolation, highlighting the value of the dense linear array.

\subsection{Baseline Model of the SFV Receiver Functions}

\begin{figure}[ht]
\centering
\includegraphics[width=\textwidth]{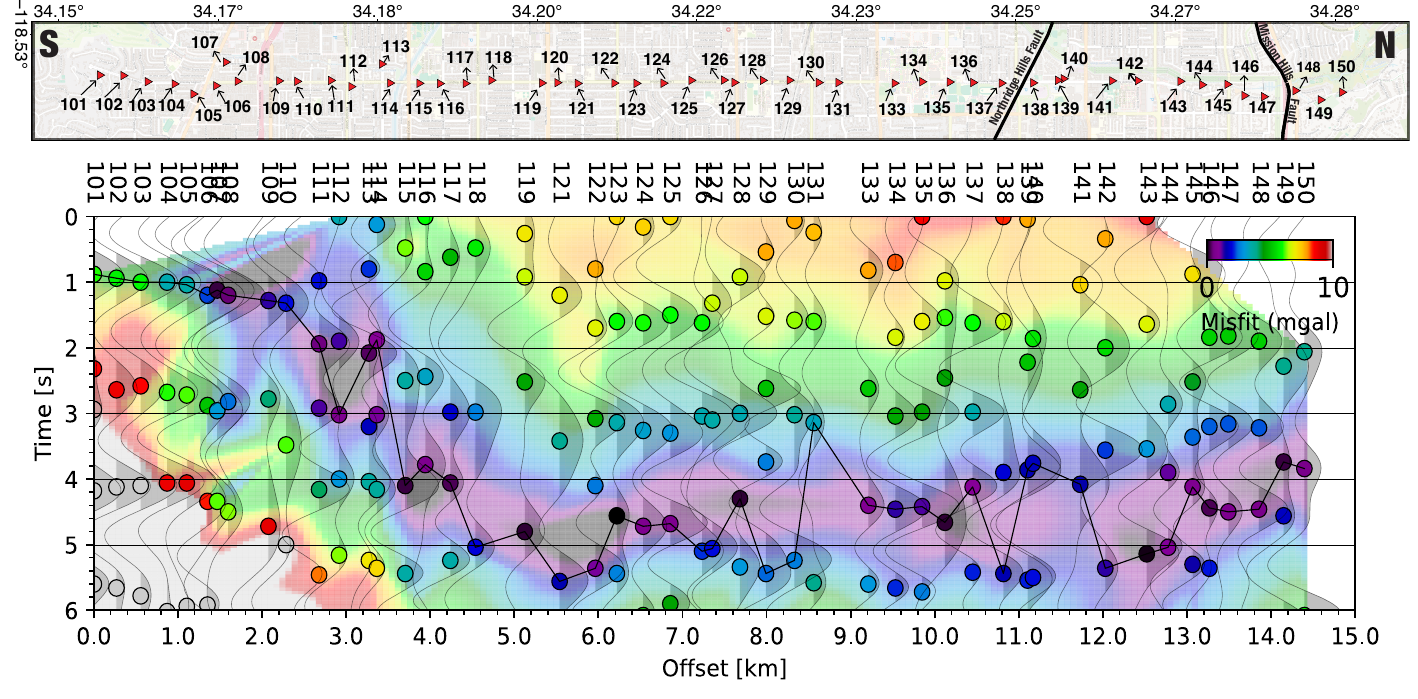}
\caption{Gravity-guided interpretations of 1 Hz receiver functions using a constant effective density contrast of -50 $kg/m^3$. Misfits associated with the possible PpPs picks are color-coded. The top panel shows the map view of station locations, with faults indicated by black lines.}
\label{grav_rfs_fig4}
\end{figure}

Line 1 yields an interpretable RF image because its dense 250 m station spacing allows coherent phases to be readily tracked across stations. The gravity-guided RFs along this line, obtained from Bouguer’s formula with an assumed constant density contrast of –50 kg/m³, highlight a potential path for identifying the sediment–basement interface using the PpPs phase (Fig. \ref{grav_rfs_fig4}).  Figure S5 shows the resulting basin depth map for the full SFV array derived using the \citeA{juarezzuniga_new_2025} velocity model. 

Along the first ten stations of Line 1, misfit values are systematically higher, reflecting the positive gravity anomaly in this segment (Fig. \ref{grav_rfs_fig4}). This suggests that the basement is exposed here, making the assumed sediment–basement density contrast of –50 kg/m³ inappropriate. As a result, the Bouguer-based prediction cannot reproduce the observations, leading to poorer fits in this region. Even so, the first pick still provides the best fit compared to deeper time pick alternatives. This example shows that while the method is effective for identifying first-order paths, areas outside the basin with positive gravity anomalies likely require additional information to guide the best pick.

Further north along Line 1, the path itself becomes ambiguous. For example, across stations 133–138, two phases emerge with similarly low misfit values, making it unclear which corresponds to the interface. In such cases, incorporating spatial coherence across neighboring stations starts to become essential for resolving the ambiguity.

\subsection{Bayesian Synthetic Test}

\begin{figure}[ht]
\centering
\includegraphics[width=\textwidth]{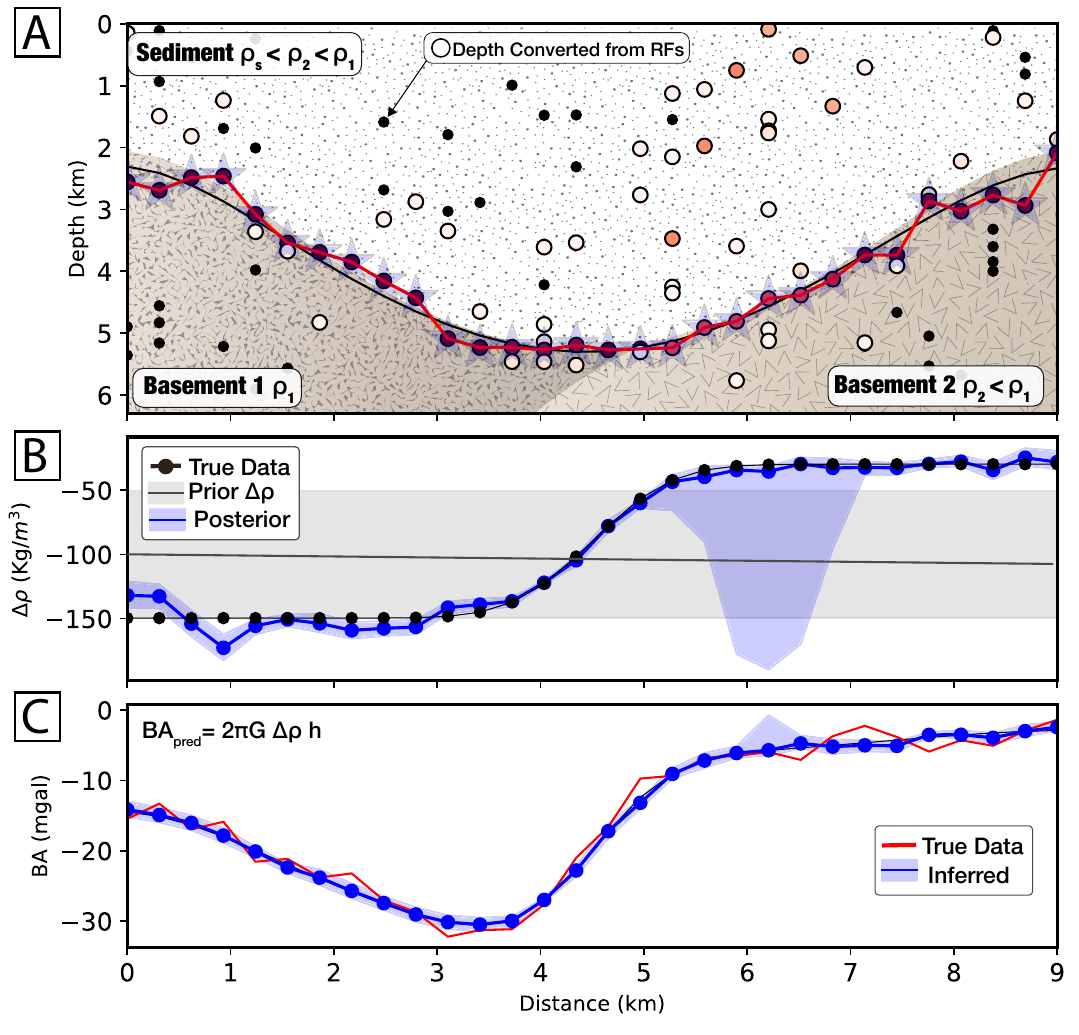}
\caption{A) GravNet synthetic results along an arbitrary profile.  The textured brown basement variants indicate different basement types; sedimentary rocks are shown in white. The points on panel (A) are colored by probability, as inferred from the posterior distribution of associated depths. Blue stars on panel (A) indicate the median time of that probability distribution, i.e., the most likely time to basement. Panel (B) shows the density posterior along with 16th and 84th percentiles. Panel (C) shows the median BA and 16th/84th percentiles.}
\label{synthetic_fig5}
\end{figure}

We present gravity-guided receiver-function picks using Bayesian inference with graph Laplacian regularization on a simple synthetic model to demonstrate the flexibility and accuracy of the method (Fig. \ref{synthetic_fig5}). This synthetic test is useful because, unlike real receiver functions, where the basement depth is uncertain, here the ground truth is known, allowing us to directly evaluate performance. We do not use simulated receiver-function times. Our goal is to test whether the Bayesian model can recover the correct density contrast and depth path from noisy depth observations. To generate the ground truth depth, we first define a sinusoidal depth curve and then perturb it with Gaussian noise. To test how the method behaves when several plausible depths are available at each station, we draw 4-7 additional depth values from a uniform window around the true depth. 

The synthetic example was designed to test the model’s ability to recover basin structure under varying density contrasts. In this setup, two different basement complexes with densities of –150 and –20 kg/m³ are overlain by the same sedimentary layer, forming a bowl-shaped sediment–basement interface (Fig. \ref{synthetic_fig5}a). To simulate the noise present in real data, we generated random time picks to represent P-to-S conversions from a ground-truth interface (black line). A linear interpolation connects the two basement densities (Fig. \ref{synthetic_fig5}b). Finally, the true Bouguer gravity response was calculated from Bouguer’s formula using the ground-truth depth and density values (Fig. \ref{synthetic_fig5}c).

The synthetic example highlights the strengths of the inference model in delineating the basin bottom under varying density contrasts. The path that maximizes the posterior is closest to the ground-truth depth (Fig. \ref{synthetic_fig5}a), as seen in the density distribution where the two true densities are accurately recovered. By contrast, assuming a constant density in this more complex case fails to guide the gravity to the correct time pick, producing a poor fit across much of the profile (Fig. S6). Similarly, removing the regularization causes the picks that maximize the posterior to scatter widely and deviate from the ground truth (Fig. S7). Even when assuming either one of the end-member densities or their average, the misfit between the ground truth and the inferred pick remains large. These outcomes illustrate the non-uniqueness of gravity: the observed value can be matched by trading off density with multiple possible depths, leading to large errors if density variability and spatial coherence are not accounted for. If we choose a very large density regularization (large $\lambda_\rho$) term, the computed density contrast becomes essentially constant across all receivers. Likewise, a very large depth regularization (large $\lambda_{depth}$) makes the inferred interface nearly flat. Therefore, the selection of the number of stations is important, as it controls how much information is shared between receivers.

\subsection{3D Bayesian Inference Results for the SFV}

\begin{figure}[ht]
\centering
\includegraphics[width=\textwidth]{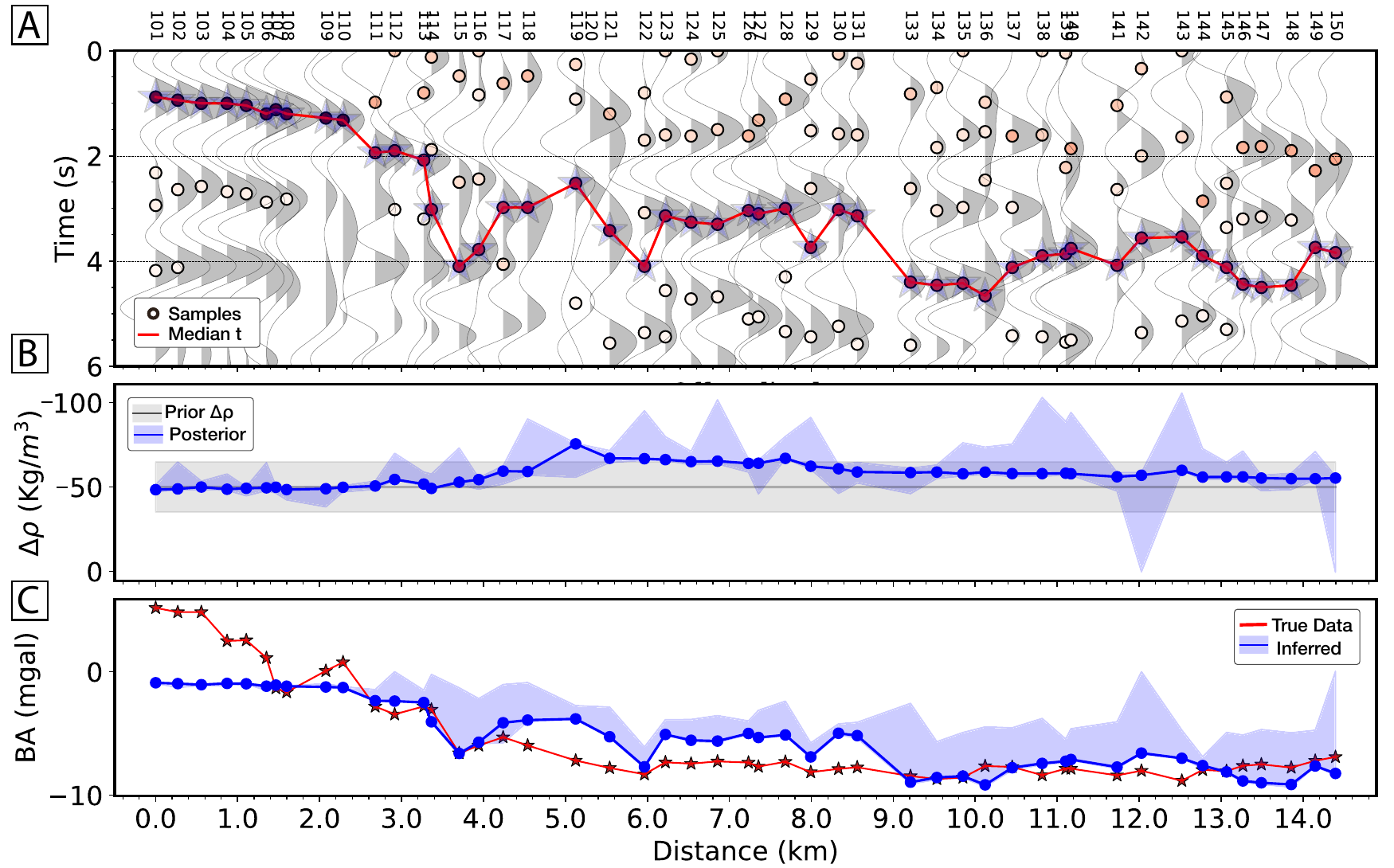}
\caption{GravNet results along Line 1, the south-to-north station profile of the SFV array. The points on panel (A) are colored by probability, as inferred from the posterior distribution of associated depths. Blue stars on panel (A) indicate the median time of that probability distribution, i.e., the most likely time to basement. Panel (B) shows the density posterior along with 16th and 84th percentiles, from which other quantities are calculated. Panel (C) shows the median BA and 16th/84th percentiles.}
\label{gravresult_fig6}
\end{figure}

We apply the 3D Bayesian inference method to the SFV data. In Figure \ref{gravresult_fig6}, we highlight the effectiveness of this method along Line 1, although it is applied to all stations, as shown by the graph network in Figure S8. Open circles represent times that were sampled by this method (we note that time is an indirectly sampled quantity, inferred from the density contrasts sampled by MCMC). The times shown are picks identified as the PpPs phase --- the predicted BA  values are calculated using the depth converted from those picks as outlined in Equation \ref{eq:PpPs}. Open circles are colored by the amount that a particular time is sampled. The red line represents the median of the generated time distribution per station. Blue stars are also present in \ref{gravresult_fig6}a to highlight these points along the red line. In the middle panel, the prior distribution of the density contrast is shown in gray. The blue dots and the connecting line are the median of the posterior distribution sampled by MCMC. The lower and upper regions represent the 16th and 84th percentiles, respectively. In the bottom panel, the red stars represent the Bouguer anomaly data of the stations. The blue dots and the connecting line are the median of the posterior distribution, though, like time, the BA is an indirectly sampled quantity inferred from the density contrast. 

We gather from Figure \ref{gravresult_fig6} that in the first ten stations at the southern end of the profile, the median density contrast remains constant and the inferred values of time are low. At this point, the stations are outside the basin, leading to poor agreement between the true and inferred BA. Nevertheless, the inferred times are consistent with a north-south upward slope as we get closer to the foothills. Moving northward along the line, the inferred values of time become greater in response to the larger median values of density contrast. At the northern end of the line, there is some evidence for the density contrasts returning to lower values. 

\begin{figure}[ht]
\centering
\includegraphics[width=\textwidth]{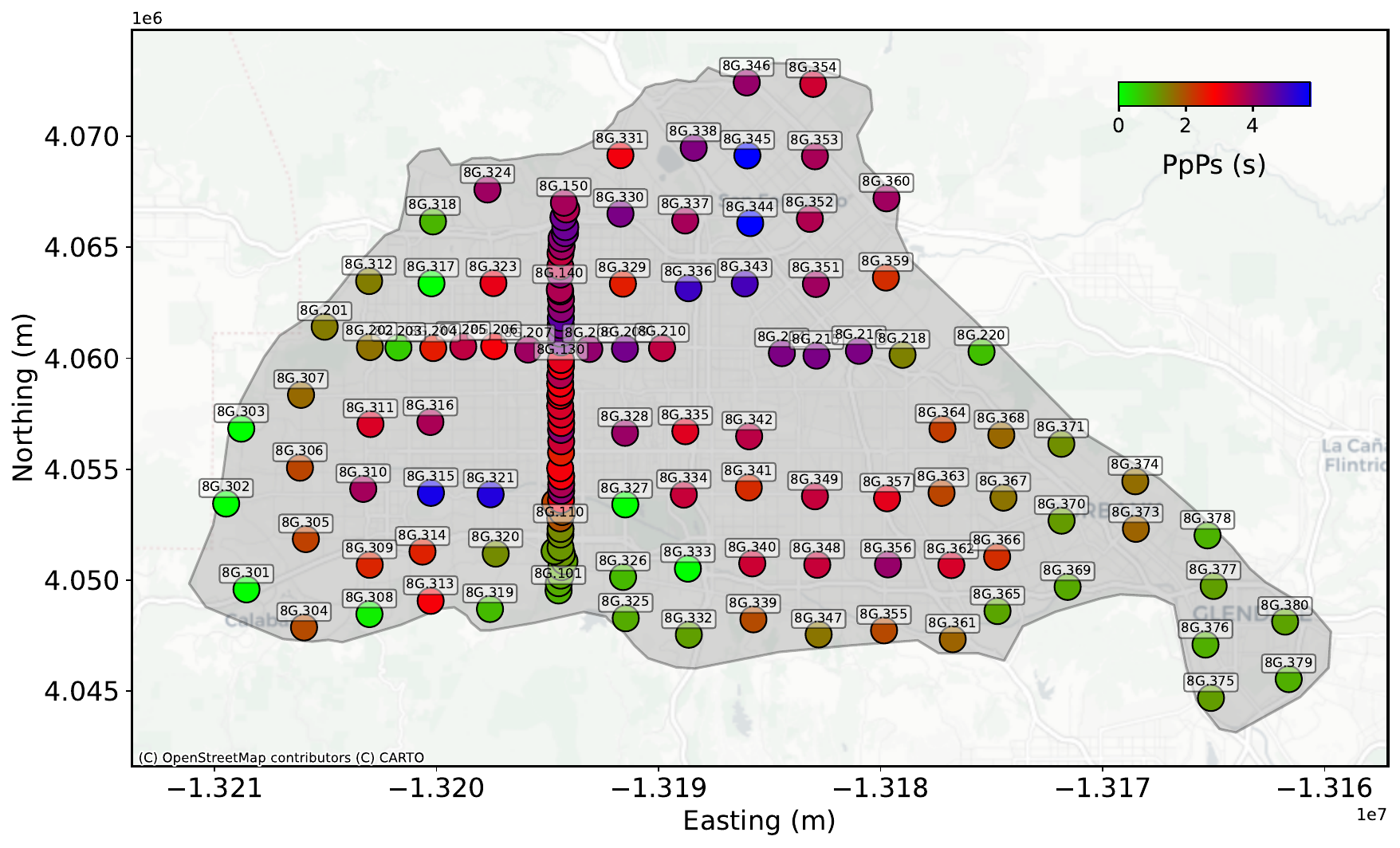}
\caption{3D Bayesian inferred time to basement in the San Fernando Valley. The time shown is identified as the PpPs time. }
\label{3dtime_fig7}
\end{figure}

Results for the 3D Bayesian inference case are shown in Figure \ref{3dtime_fig7}. The south-to-north line is Line 1 as referred to above. Along the southern edge, the times are the lowest, indicating the shallowest part of the basin. In the San Fernando sub-basin (see Figure 1 for location), the values of PpPs times are larger reaching values greater than 4s. This is also seen in the Sylmar sub-basin. In the southeast part of the San Fernando Valley, this pattern is not observed, and the PpPs time shows less variability throughout. The median density inferred shows a range of effective density contrasts between -22 and -75 $kg/m^3$, and with the exception of the ones at the edge of the basin, have a low misfit (Fig. S9). Upon comparison with the result of \citeA{juarezzuniga_new_2025}, changes in the effective density contrast along Line 1 appear to reflect the faults that produce offsets in the basement. For example, the pattern north of Station 110 in Figure S9 appears to coincide with the Leadwell fault and the changes between Stations 130 and 140 appear to trace the Northridge Hills fault. As outlined in Figure 5 using synthetic data, such contrasts in rock properties is exactly what our method is intended to detect --- changes in density due to different basement types.

\begin{figure}[ht]
\centering
\includegraphics[width=\textwidth]{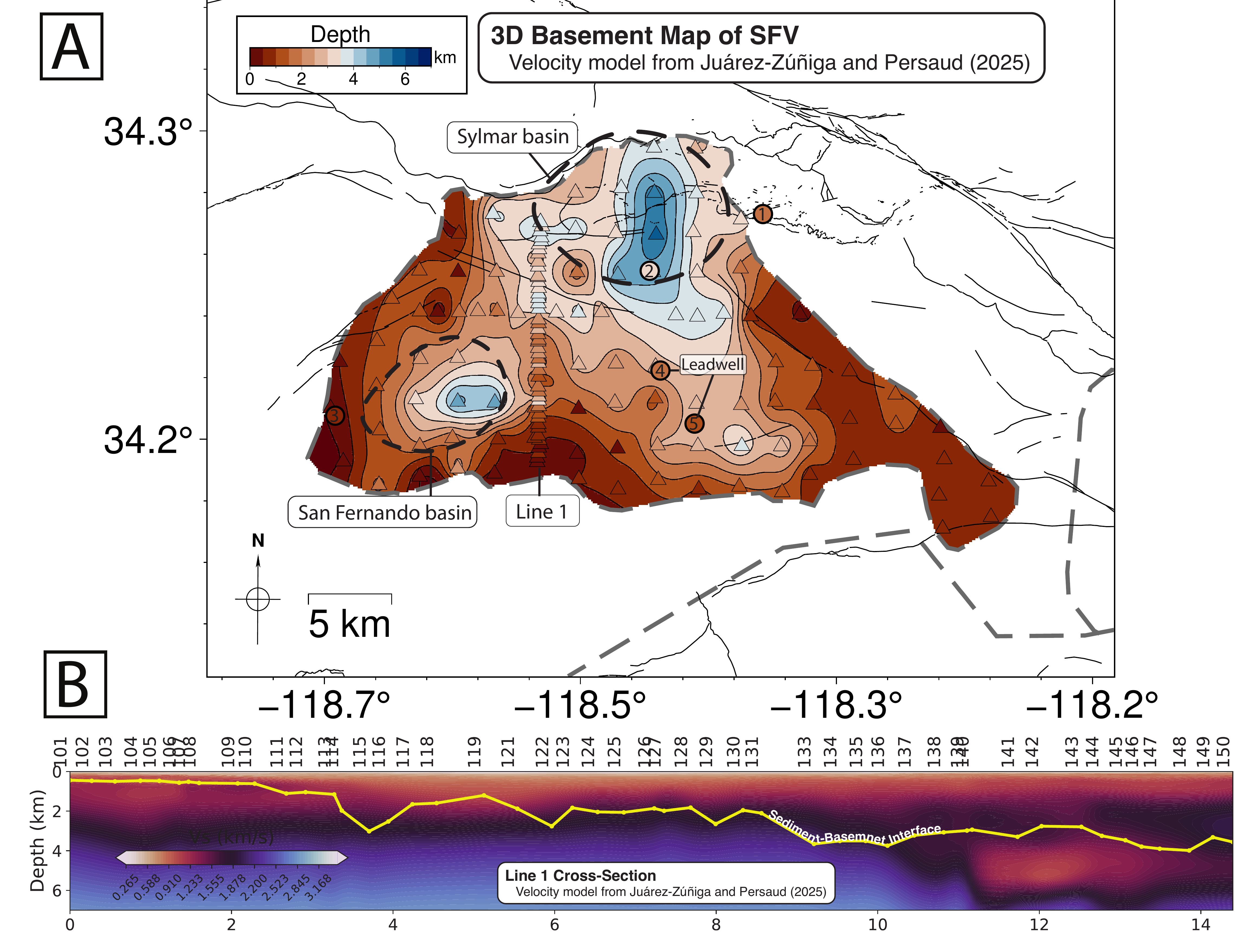}
\caption{(A) 3D Bayesian Inference Depth model for the San Fernando Valley using the velocity model of \citeA{juarezzuniga_new_2025}. (B) Shear-wave velocity cross-sections along Line 1 extracted from the velocity model. The yellow lines show the corresponding basin depths.}
\label{depthmap_fig8}
\end{figure}

In Figure \ref{depthmap_fig8}, the inferred depths agree with the qualitative analysis of times in the previous paragraph. In the case of applying the \citeA{juarezzuniga_new_2025} velocity model, the depth of the Sylmar sub-basin reaches 5.6 km, exceeding that of the San Fernando sub-basin. In addition, the southern-central part is shallow and matches the two boreholes in the area (numbered 4 and 5, Table S1).

\begin{figure}[ht]
\centering
\includegraphics[width=\textwidth]{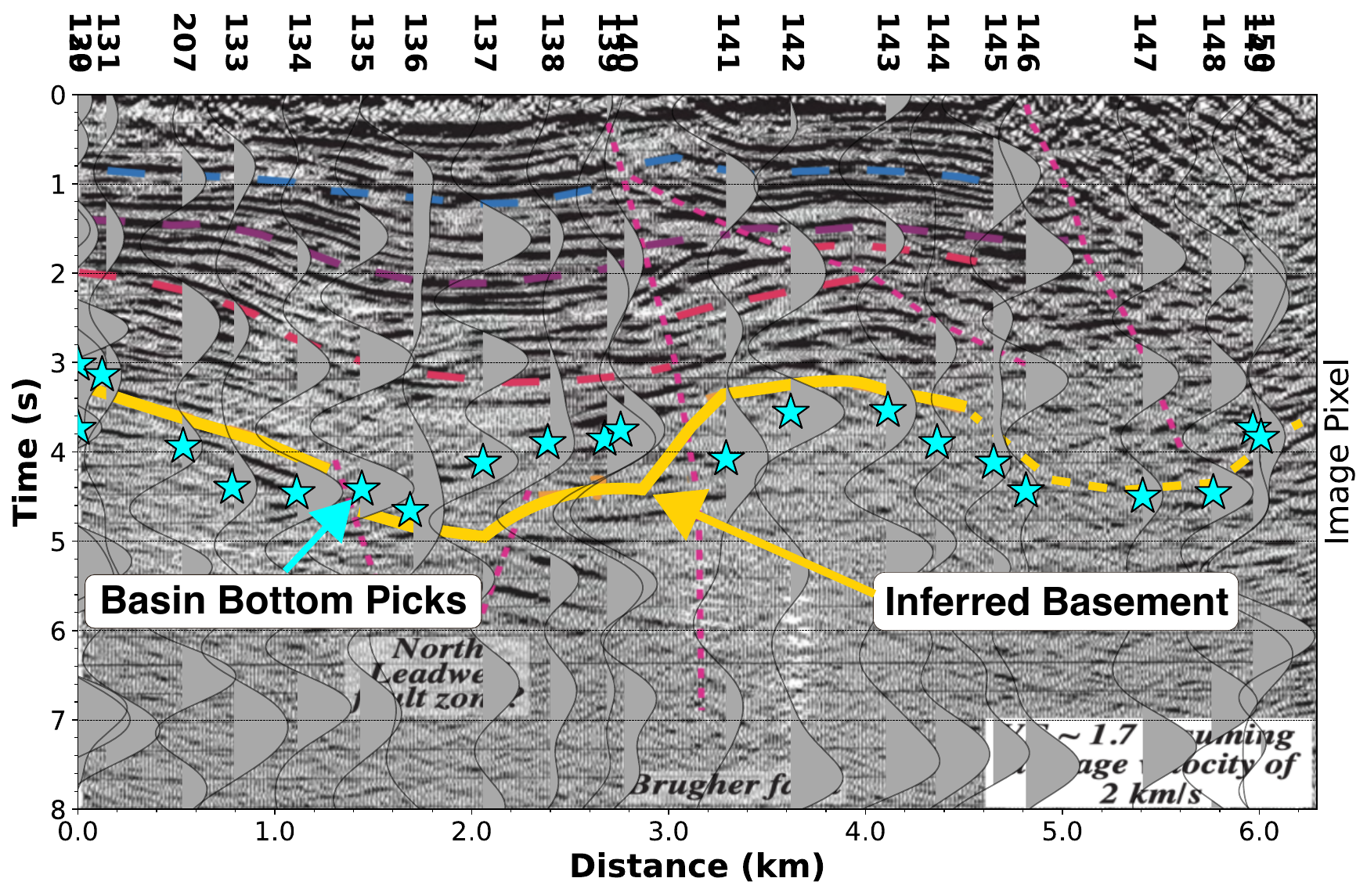}
\caption{Receiver Functions overlaid on industry profiles SFV-85-11 \cite{langenheim_structure_2011}. The stars indicate the best pick for the sediment-basement interface. }
\label{industryprofile_fig9}
\end{figure}

Figure \ref{industryprofile_fig9} shows that the PpPs times inferred from the 3D Bayesian model are in good agreement with the industry profile interpretation of the basement along Line 1 \cite{langenheim_structure_2011}. The median times from the RFs PpPs phases trace the inferred basement line imaged in the industry reflection profile, outlining the same coherent trend. A visual inspection indicates that other phases could plausibly be interpreted as additional layers; however, the layer inferred by the 3D Bayesian model aligns with the industry near-basement interpretation. A similar level of agreement is also observed in industry profiles 2893-O and SFV-85-12 (Fig. S10-11).

\section{Discussion}

This work introduces a new method for processing RFs in urban seismology. The 3D Bayesian inference technique quantitatively arrives at the best density contrast, which in turn yields the best time picks corresponding to the sedimentary basin bottom. The success of this technique lies in the nearest neighbor approach, where the model favors picks based on trends of the nearest neighbors. Regularization controls how much weight each neighbor imparts. Integrating this into the likelihood and combining with informed priors on the density contrast yields a fully Bayesian probabilistic model. 

The method applies to arbitrary station geometry and allows us to arrive at a single best solution. The Bouguer Anomaly fit helps highlight potential limitations: when the signal is dominated by noise or incoherency, a large misfit indicates that the selected path is unlikely to be optimal. Although the model generally identifies the most coherent path, even with variable data quality or station density, such cases call for additional constraints or independent validation to ensure reliable interpretation. Another important consideration is that the Bouguer anomaly fit constrains an effective density contrast that reflects the spatial scales emphasized by the gravity field and filtering applied, rather than a unique lithologic density contrast.

\subsection{Geological Interpretation}
The analysis of Line 1 along the industry profiles illustrates the complexity of the SFV. The profiles SFV-85-11 and 2893-O show the following important layers: the top of the Sunshine Ranch (blue), the base of the Saugus (purple), the top of the Modelo (pink), and the top of the Topanga (orange) (dashed lines in Fig. \ref{industryprofile_fig9}, Fig. S10). Our results align with the top of the Topanga layer, which reflects Middle Miocene volcanic rocks. Beneath the Topanga formation are the pre-San Fernando basin formation. These include the Sespe, Llajas, Chatsworth formations and quartz diorite, gneisses rock, and Santa Monica slate as basement rocks \cite{langenheim_structure_2011}. Profile SFV-85-12 also shows this Topanga-basement contact reflection (Fig. S11). Since the Topanga Formation rests directly on older sedimentary units or crystalline basement, we interpret the gravity-guided RF picks as marking the closest resolvable contact to the basement. Along some segments of the profile, the density contrast differs from that of neighboring stations, pointing to local variations in the geology. In the region, for example, the Topanga Formation is known to thin and eventually wedge out northward onto the crest of the Leadwell high \cite{langenheim_structure_2011}.

Our comparison with borehole control on the Leadwell High shows that the velocity model of \cite{juarezzuniga_new_2025} provides a shallow depth structure consistent with ground truth. Two boreholes (3 and 4) confirm the presence of basement in this region (Fig. \ref{3dtime_fig7}, Table S1). The Leadwell High is a concealed basement high in the southeastern SFV, primarily composed of granitic rocks. Our depth model, derived from the \cite{juarezzuniga_new_2025} velocities, matches the shallow borehole depths.

Although not as well documented in boreholes, our results show consistencies with the syncline structure mapped by \citeA{langenheim_structure_2011} west of the line, a feature that is often overlooked. A similar depth structure appears in the velocity model of \citeA{juarezzuniga_new_2025}, with our model placing it slightly farther east. This region is characterized by a pronounced angular unconformity at the base of the Modelo Formation and strong reflections above, interpreted as Tarzana fan deposits \cite{langenheim_structure_2011}. By reproducing this feature, our results strengthen the evidence for this sub-basin and support the interpretation that it is largely filled by Tarzana fan deposits.

Recent studies, including ours, place the deepest part of the Sylmar basin between the Mission Hills fault to the south and the San Fernando fault to the north \cite{juarezzuniga_new_2025}, in contrast to the geometry mapped by \citeA{langenheim_structure_2011}, who bounded the basin farther north between the Mission Hills and Santa Susana—Hospital–Sierra Madre fault system and reported depths of 5–8 km. Such enhancements in the basin geometry and near-fault rock properties can result in substantial changes in ground shaking estimates for the San Fernando Valley. Our model yields a maximum depth of 5.6 km, slightly shallower than the 6 km reported by \citeA{juarezzuniga_new_2025}, whose velocity model we adopt. The discrepancy with earlier work likely reflects (i) differences in the treatment of residual Bouguer gravity, which all studies use to shape basin geometry, and (ii) limited seismic-reflection and well control in the northern area emphasized by \citeA{langenheim_structure_2011}, where gravity alone can exaggerate depths.

Our Bayesian formulation also differs from the gravity-based approach of \citeA{juarezzuniga_new_2025} in how density contrast and basin depth are incorporated. In \citeA{juarezzuniga_new_2025}, the shear-wave velocity model was first converted to density, and the resulting density structure was used to invert gravity and interpret HVSRs to map the basin bottom in more detail; the density field was therefore fixed by the velocity model, and basin depth was adjusted to match the observed Bouguer anomaly. In the present work, we use the same velocity model primarily to convert receiver-function information into depth and to define a small set of plausible depth candidates at each station, which we treat as discrete observations rather than a continuous model field. We do not prescribe density contrast as a deterministic function of velocity; instead, the density contrasts are treated as unknown parameters with explicit priors and graph-based regularization, and the model evaluates which depth candidates are most consistent with those densities and with neighboring stations. This makes the tradeoffs between depth and density an explicit part of the inference. For example, a smaller density contrast can often be compensated by a shallower interface, whereas a larger contrast permits deeper solutions that still fit the same depth picks. The regularization parameters $\lambda_\rho$, $\lambda_h$, and $k$ control how strongly such local tradeoffs are tied together across the array. Because the method can highlight any sufficiently strong impedance contrast, prior geological knowledge about which interface is expected in a given depth range remains essential when interpreting the recovered structure.

\section{Conclusion}
We present a tool to quantitatively tackle the ambiguity of interpreting dense array RFs to obtain sedimentary basin structure and layer properties, particularly in urban RF studies where such constraints are critical for improving seismic hazard estimates. Our method guides RF picks using gravity measurements with a flexible density contrast parameter. Such an approach has received virtually no attention in the past, partly due to the lack of dense urban array datasets. The 3D Bayesian inference model assigns different density contrasts to each neighborhood based on the best coherence of phase picks in the receiver function. The regularization sets how much weight neighboring stations provide when sampling the best density contrast, and effectively, the time picked. This process gives a quantitative way to identify the sediment–basement interface.

The power of this tool is demonstrated in the newly acquired 2023 SFV dense nodal array dataset that was recoded in a densely populated urban area \cite{patricia_persaud_san_2023}. The 3D Bayesian inference time-to-basement map and the depth map from the more detailed velocity model of \citeA{juarezzuniga_new_2025} show the most prominent geologic features such as the Leadwell High, the San Fernando sub-basin, and the Sylmar sub-basin. The San Fernando sub-basin reaches a maximum depth of 4 km, and the Sylmar basin 5.6 km. With urban seismology entering a new era of dense deployments, tools like this will be crucial for identifying key layers such as the sediment–basement interface.

\section{Open Research}

The Graph-Regularized Bayesian Approach for Gravity-Informed Mapping (GravNet) software is publicly available at \citeA{villa_graph-regularized_2025}. The time to basement and depth values from this study (along with the software) are publicly available at \\\url{https://github.com/vvillaga/GravNet}.

The basement depths obtained from borehole logs are publicly available through the California Geologic Energy Management Division's (CalGEM) online mapping application Well Finder. The Bouguer gravity data were provided by the Pan American Center Earth and Environmental Science portal \cite{paces_paces_2012}. The portal is no longer active, but the dataset is available at \url{http://dx.doi.org/10.22002/D1.20256} \cite{clayton_gravity_2022}. Seismic data from the San Fernando Valley (SFV) Nodal Array, network code 8G \cite{patricia_persaud_san_2023} were downloaded through the EarthScope Consortium Web Services (https://service.iris.edu/, last accessed April 2025). The facilities of EarthScope Consortium were used for access to waveforms and related metadata. Figures were plotted using the GMT software, PyGMT, and Cartopy \cite{met_office_cartopy_2010, wessel_generic_2019, tian_pygmt_2025}.


\acknowledgments
The authors thank the San Fernando Valley (SFV) residents who hosted the seismic stations and members of the SFV Nodal Array installation and pickup teams. EarthScope provided the seismic instruments through the EarthScope Primary Instrument Center (EPIC, formerly the Program for the Array Seismic Studies of the Continental Lithosphere, PASSCAL Instrument Center) at New Mexico Tech. This work has been supported by the Department of Geosciences at the University of Arizona, the U.S. Geological Survey (USGS) Award Number G24AP00067, the Statewide California Earthquake Center Award Number 24148, and the National Science Foundation (NSF) Award Numbers 2105320, 2105358, and 2438773. The facilities of the EarthScope Consortium are supported by the NSF’s Seismological Facilities for the Advancement of Geoscience (SAGE) Award under Cooperative Support Agreement EAR-1851048. V.V. acknowledges support from the National Science Foundation through a Graduate Research Fellowship.



%
\bibliography{references}

@article{piana_agostinetti_sedimentary_2019,
	title = {Sedimentary basins investigation using teleseismic {P}-wave time delays},
	volume = {67},
	issn = {1365-2478},
	doi = {10.1111/1365-2478.12747},
	language = {en},
	number = {6},
	journal = {Geophysical Prospecting},
	author = {Piana Agostinetti, Nicola and Martini, Francesca},
	year = {2019},
	pages = {1676--1685},
}

@article{persaud_volunteer-led_2024,
	title = {Volunteer-{Led}, {Short}-{Term}, {Geophysical} {Field} {Experiment}: {Lessons} for {Inviting} {Broader} {Participation}, {Building} {Public} {Trust}, and {Communicating} {Science}},
	volume = {34},
	shorttitle = {Volunteer-{Led}, {Short}-{Term}, {Geophysical} {Field} {Experiment}},
	doi = {10.1130/GSATG590GW.1},
	language = {en},
	number = {10},
	journal = {GSA Today},
	author = {Persaud, P.},
	year = {2024},
}

@article{esteve_mapping_2025,
	title = {Mapping {Basin} {Interfaces} {Using} {Single}-{Station} {Cross}-{Component} {Correlations}: {Application} to the {Central} {Vienna} {Basin} ({Austria})},
	volume = {52},
	copyright = {© 2025. The Author(s).},
	issn = {1944-8007},
	shorttitle = {Mapping {Basin} {Interfaces} {Using} {Single}-{Station} {Cross}-{Component} {Correlations}},
	doi = {10.1029/2025GL116888},
	language = {en},
	number = {20},
	journal = {Geophysical Research Letters},
	author = {Esteve, C. and Lu, Y. and Bokelmann, G.},
	year = {2025},
}

@misc{patricia_persaud_san_2023,
	title = {San {Fernando} {Valley}, {California} {Nodal} {Array} (2023)},
	url = {https://www.fdsn.org/networks/detail/8G_2023/},
	doi = {10.7914/7XH3-5A25},
	publisher = {International Federation of Digital Seismograph Networks},
	author = {Patricia Persaud},
	year = {2023},
}

@article{ramirez_moho_2021,
	title = {Moho {Depth} of {Northern} {Baja} {California}, {Mexico}, {From} {Teleseismic} {Receiver} {Functions}},
	volume = {8},
	copyright = {© 2021. The Authors.},
	issn = {2333-5084},
	doi = {10.1029/2020EA001463},
	language = {en},
	number = {6},
	journal = {Earth and Space Science},
	author = {Ramírez, E. E. and Bataille, Klaus and Vidal-Villegas, J. A. and Stock, J. M. and Ramírez-Hernández, J.},
	year = {2021},
}

@article{lawrence_global_2006,
	title = {A global study of transition zone thickness using receiver functions},
	volume = {111},
	issn = {2156-2202},
	url = {https://onlinelibrary.wiley.com/doi/abs/10.1029/2005JB003973},
	doi = {10.1029/2005JB003973},
	language = {en},
	number = {B6},
	journal = {Journal of Geophysical Research: Solid Earth},
	author = {Lawrence, Jesse F. and Shearer, Peter M.},
	year = {2006},
}

@article{palmer_san_1971,
	title = {San {Fernando} {Earthquake} of 9 {February} 1971: {Pattern} of {Faulting}},
	volume = {172},
	shorttitle = {San {Fernando} {Earthquake} of 9 {February} 1971},
	doi = {10.1126/science.172.3984.712},
	number = {3984},
	journal = {Science},
	author = {Palmer, D. F. and Henyey, T. L.},
	year = {1971},
	pages = {712--715},
}

@article{hough_17_2024,
	title = {The 17 {January} 1994 {Northridge}, {California}, {Earthquake}: {A} {Retrospective} {Analysis}},
	volume = {4},
	issn = {2694-4006},
	shorttitle = {The 17 {January} 1994 {Northridge}, {California}, {Earthquake}},
	doi = {10.1785/0320240012},
	number = {3},
	journal = {The Seismic Record},
	author = {Hough, Susan E. and Graves, Robert W. and Cochran, Elizabeth S. and Yoon, Clara E. and Blair, Luke and Haefner, Scott and Wald, David J. and Quitoriano, Vincent},
	year = {2024},
	pages = {151--160},
}

@article{yan_regional_2007,
	title = {Regional mapping of the crustal structure in southern {California} from receiver functions},
	volume = {112},
	issn = {0148-0227},
	url = {http://doi.wiley.com/10.1029/2006JB004622},
	doi = {10.1029/2006JB004622},
	language = {en},
	number = {B5},
	journal = {Journal of Geophysical Research},
	author = {Yan, Z. and Clayton, R. W.},
	year = {2007},
	pages = {B05311},
}

@article{wessel_generic_2019,
	title = {The {Generic} {Mapping} {Tools} {Version} 6},
	volume = {20},
	issn = {1525-2027},
	doi = {10.1029/2019GC008515},
	language = {en},
	number = {11},
	journal = {Geochemistry, Geophysics, Geosystems},
	author = {Wessel, P. and Luis, J. F. and Uieda, L. and Scharroo, R. and Wobbe, F. and Smith, W. H. F. and Tian, D.},
	year = {2019},
	pages = {5556--5564},
}

@article{wang_urban_2021,
	title = {Urban {Basin} {Structure} {Imaging} {Based} on {Dense} {Arrays} and {Bayesian} {Array}-{Based} {Coherent} {Receiver} {Functions}},
	volume = {126},
	issn = {2169-9356},
	doi = {10.1029/2021JB022279},
	language = {en},
	number = {9},
	journal = {Journal of Geophysical Research: Solid Earth},
	author = {Wang, Xin and Zhan, Zhongwen and Zhong, Minyan and Persaud, Patricia and Clayton, Robert W.},
	year = {2021},
}

@article{villa_three-dimensional_2023,
	title = {Three-{Dimensional} {Basin} {Depth} {Map} of the {Northern} {Los} {Angeles} {Basins} {From} {Gravity} and {Seismic} {Measurements}},
	volume = {128},
	copyright = {© 2023. American Geophysical Union. All Rights Reserved.},
	issn = {2169-9356},
	doi = {10.1029/2022JB025425},
	language = {en},
	number = {7},
	journal = {Journal of Geophysical Research: Solid Earth},
	author = {Villa, Valeria and Li, Yida and Clayton, Robert W. and Persaud, Patricia},
	year = {2023},
	pages = {e2022JB025425},
}

@article{bard_two-dimensional_1985,
	title = {The two-dimensional resonance of sediment-filled valleys},
	volume = {75},
	issn = {0037-1106},
	doi = {10.1785/BSSA0750020519},
	number = {2},
	journal = {Bulletin of the Seismological Society of America},
	author = {Bard, Pierre-Yves and Bouchon, Michel},
	year = {1985},
	pages = {519--541},
}

@article{juarezzuniga_new_2025,
	title = {New {Insights} into the {Crustal} {Structure} of the {San} {Fernando} {Valley}, {California}, from a {Dense} {Nodal} {Seismic} {Array}},
	issn = {0895-0695},
	doi = {10.1785/0220240473},
	journal = {Seismological Research Letters},
	author = {Juárez‐Zúñiga, Alan and Persaud, Patricia},
	year = {2025},
}

@misc{clayton_gravity_2022,
	title = {Gravity {Data} {For} {Southern} {California}},
	publisher = {CaltechDATA},
	author = {Clayton, Robert},
	year = {2022},
	doi = {10.22002/D1.20256},
}

@article{zhu_moho_2000,
	title = {Moho depth variation in southern {California} from teleseismic receiver functions},
	volume = {105},
	issn = {2156-2202},
	doi = {10.1029/1999JB900322},
	language = {en},
	number = {B2},
	journal = {Journal of Geophysical Research: Solid Earth},
	author = {Zhu, Lupei and Kanamori, Hiroo},
	year = {2000},
	pages = {2969--2980},
}

@article{vinnik_detection_1977,
	title = {Detection of waves converted from {P} to {SV} in the mantle},
	volume = {15},
	issn = {0031-9201},
	doi = {10.1016/0031-9201(77)90008-5},
	number = {1},
	journal = {Physics of the Earth and Planetary Interiors},
	author = {Vinnik, L. P.},
	year = {1977},
	pages = {39--45},
}

@misc{villa_graph-regularized_2025,
	title = {Graph-{Regularized} {Bayesian} {Approach} for {Gravity}- {Informed} {Mapping} ({GravNet})},
	publisher = {Zenodo},
	author = {Villa, Valeria},
	year = {2025},
	note = {doi 10.5281/zenodo.17728056},
}

@misc{us_census_bureau_san_nodate,
	title = {San {Fernando} {Valley} {CCD}, {Los} {Angeles} {County}, {California} ({Geographic} {Profile})},
	author = {{U.S. Census Bureau}},
}

@misc{tian_pygmt_2025,
	title = {{PyGMT}: {A} {Python} interface for the {Generic} {Mapping} {Tools}},
	publisher = {Zenodo},
	author = {Tian, Dongdong and Uieda, Leonardo and Leong, Wei Ji and Fröhlich, Yvonne and Grund, Michael and Schlitzer, William and Jones, Max and Toney, Liam and Yao, Jiayuan and Tong, Jing-Hui and Magen, Yohai and Materna, Kathryn and Belem, Andre and Newton, Tyler and Anant, Abhishek and Ziebarth, Malte and Quinn, Jamie and Wessel, Paul},
	year = {2025},
}

@article{ozakin_systematic_2015,
	title = {Systematic {Receiver} {Function} {Analysis} of the {Moho} {Geometry} in the {Southern} {California} {Plate}-{Boundary} {Region}},
	volume = {172},
	issn = {1420-9136},
	doi = {10.1007/s00024-014-0924-6},
	language = {en},
	number = {5},
	journal = {Pure and Applied Geophysics},
	author = {Ozakin, Yaman and Ben-Zion, Yehuda},
	year = {2015},
	pages = {1167--1184},
}

@article{omojola_detecting_2025,
	title = {Detecting {Urban} {Earthquakes} with the {San} {Fernando} {Valley} {Nodal} {Array} and {Machine} {Learning}},
	issn = {0895-0695},
	doi = {10.1785/0220250124},
	journal = {Seismological Research Letters},
	author = {Omojola, Joses and Persaud, Patricia},
	year = {2025},
}

@book{met_office_cartopy_2010,
	address = {Exeter, Devon},
	title = {Cartopy: a cartographic python library with a matplotlib interface},
	author = {{Met Office}},
	year = {2010},
}

@article{ma_structure_2016,
	title = {Structure of the {Los} {Angeles} {Basin} from ambient noise and receiver functions},
	volume = {206},
	issn = {0956-540X},
	doi = {10.1093/gji/ggw236},
	number = {3},
	journal = {Geophysical Journal International},
	author = {Ma, Yiran and Clayton, Robert W.},
	year = {2016},
	pages = {1645--1651},
}

@article{liu_structure_2018,
	title = {Structure of the {Northern} {Los} {Angeles} {Basins} {Revealed} in {Teleseismic} {Receiver} {Functions} from {Short}‐{Term} {Nodal} {Seismic} {Arrays}},
	volume = {89},
	issn = {0895-0695, 1938-2057},
	doi = {10.1785/0220180071},
	language = {en},
	number = {5},
	journal = {Seismological Research Letters},
	author = {Liu, Guibao and Persaud, Patricia and Clayton, Robert W.},
	year = {2018},
	pages = {1680--1689},
}

@article{langenheim_structure_2011,
	title = {Structure of the {San} {Fernando} {Valley} region, {California}: {Implications} for seismic hazard and tectonic history},
	volume = {7},
	issn = {1553-040X},
	shorttitle = {Structure of the {San} {Fernando} {Valley} region, {California}},
	doi = {10.1130/GES00597.1},
	number = {2},
	journal = {Geosphere},
	author = {Langenheim, Victoria E. and Wright, T.L. and Okaya, D.A. and Yeats, R.S. and Fuis, G.S. and Thygesen, K. and Thybo, H.},
	year = {2011},
	pages = {528--572},
}

@article{foreman-mackey_emcee_2019,
	title = {emcee v3: {A} {Python} ensemble sampling toolkit for affine-invariant {MCMC}},
	volume = {4},
	doi = {10.21105/joss.01864},
	number = {43},
	journal = {The Journal of Open Source Software},
	author = {Foreman-Mackey, Daniel and Farr, Will and Sinha, Manodeep and Archibald, Anne and Hogg, David and Sanders, Jeremy and Zuntz, Joe and Williams, Peter and Nelson, Andrew and de Val-Borro, Miguel and Erhardt, Tobias and Pashchenko, Ilya and Pla, Oriol},
	year = {2019},
	pages = {1864},
}

@article{clayton_exposing_2019,
	title = {Exposing {Los} {Angeles}’s {Shaky} {Geologic} {Underbelly}},
	volume = {100},
	issn = {2324-9250},
	doi = {10.1029/2019EO135099},
	language = {en},
	journal = {Eos},
	author = {Clayton, Robert and Persaud, Patricia and Denolle, Marine and Polet, Jascha},
	year = {2019},
}

@article{bonilla_site_1997,
	title = {Site amplification in the {San} {Fernando} {Valley}, {California}: {Variability} of site-effect estimation using the {S}-wave, coda, and {H}/{V} methods},
	volume = {87},
	issn = {0037-1106},
	shorttitle = {Site amplification in the {San} {Fernando} {Valley}, {California}},
	doi = {10.1785/BSSA0870030710},
	number = {3},
	journal = {Bulletin of the Seismological Society of America},
	author = {Bonilla, Luis Fabián and Steidl, Jamison H. and Lindley, Grant T. and Tumarkin, Alexei G. and Archuleta, Ralph J.},
	year = {1997},
	pages = {710--730},
}

@techreport{langenheim_preliminary_2000,
	address = {Reston, VA},
	type = {Report},
	title = {Preliminary potential-field constraints on the geometry of the {San} {Fernando} basin, {Southern} {California}},
	language = {English},
	number = {2000-219},
	author = {Langenheim, Victoria E. and Griscom, Andrew and Jachens, R.C. and Hildenbrand, T.G.},
	year = {2000},
	note = {doi: 10.3133/ofr00219},
	pages = {39},
}

@misc{jennings_fault_2010,
	edition = {Version 2.0},
	title = {Fault {Activity} {Map} of {California}},
	publisher = {Department of Conservation, California Geological Survey},
	author = {Jennings, Charles W. and Bryant, William A.},
	year = {2010},
}

@article{ghose_basin_2023,
	title = {Basin {Structure} for {Earthquake} {Ground} {Motion} {Estimates} in {Urban} {Los} {Angeles} {Mapped} with {Nodal} {Receiver} {Functions}},
	volume = {13},
	copyright = {http://creativecommons.org/licenses/by/3.0/},
	issn = {2076-3263},
	doi = {10.3390/geosciences13110320},
	language = {en},
	number = {11},
	journal = {Geosciences},
	author = {Ghose, Ritu and Persaud, Patricia and Clayton, Robert W.},
	year = {2023},
	pages = {320},
}

@article{ligorria_iterative_1999,
	title = {Iterative deconvolution and receiver-function estimation},
	volume = {89},
	issn = {0037-1106},
	url = {https://doi.org/10.1785/BSSA0890051395},
	doi = {10.1785/BSSA0890051395},
	number = {5},
	urldate = {2023-10-26},
	journal = {Bulletin of the Seismological Society of America},
	author = {Ligorría, Juan Pablo and Ammon, Charles J.},
	month = oct,
	year = {1999},
	pages = {1395--1400},
}

@misc{paces_paces_2012,
	title = {{PACES}},
	url = {http://gis.utep.edu/paces/PACES%20Gravity%20Magnetics.htm},
	urldate = {2012-04-04},
	journal = {Pan-American Center for Earth and Environmental Studies},
	collaborator = {PACES},
	year = {2012},
}
%




%
%
%
%
%

\end{document}